\documentclass{IEEEtaes}

\usepackage{amsmath,amssymb,amsfonts}
\usepackage[ruled,linesnumbered]{algorithm2e}
\usepackage{array}
\usepackage[caption=false,font=footnotesize,labelfont=rm,textfont=rm]{subfig}
\usepackage{textcomp}
\usepackage{stfloats}
\usepackage{multirow}
\usepackage{makecell}
\usepackage{url}
\usepackage{verbatim}
\usepackage{graphicx}
\usepackage{cite}
\usepackage{hyperref}
\usepackage{booktabs}
\usepackage{wrapfig}
\usepackage{threeparttable}
\usepackage{bbding}
\usepackage{diagbox}

\hypersetup{hypertex=true,
            colorlinks=true,
            linkcolor=blue,
            anchorcolor=blue,
            citecolor=blue}
\jvol{XX}
\jnum{XX}
\jmonth{XXXXX}
\paper{1234567}
\pubyear{2022}
\doiinfo{TAES.2022.Doi Number}

\begin{document}

\title{DGAR: A Unified Domain Generalization Framework for RF-Based Human Activity Recognition}{Title for citation}

\author{Junshuo Liu}
\member{Graduate Student Member, IEEE}
\affil{Huazhong University of Science and Technology, Wuhan, China} 

\author{Xin Shi}
\member{Member, IEEE}
\affil{Tsinghua University, Beijing, China}

\author{Yunchuan Zhang}
\member{Member, IEEE}
\affil{King's College London, London, UK}

\author{Yinhao Ge}
\affil{Wuhan Second Ship Design and Research Institute, Wuhan, China} 

\author{Robert C. Qiu}
\member{Fellow, IEEE}
\affil{Huazhong University of Science and Technology, Wuhan, China}

%% \author{FOURTH D. AUTHOR}
%% \affil{University of Colorado, Colorado, USA}

\receiveddate{Manuscript received XXXXX 00, 0000; revised XXXXX 00, 0000; accepted XXXXX 00, 0000.\\
This work was supported in part by the Nation Natural Science Foundation of China under Grant No.12141107 and in part by the Interdisciplinary Research Program of HUST, 2023JCYJ012. \textit{(Corresponding author: Xin Shi.)}}

\authoraddress{Junshuo Liu and Robert C. Qiu are with the School of Electronic Information and Communications, Huazhong University of Science and Technology, Wuhan, 430074, China (email: junshuo\_liu; caiming@hust.edu.cn). Xin Shi is with the Energy Internet Research Institute, Tsinghua University, Beijing, 100085, China (email: xinshi\_bjcy@163.com). Yunchuan Zhang is with the Department of Engineering, King's College London, London, WC2R 2LS, UK (email: yunchuan.zhang@kcl.ac.uk). Yinhao Ge is with Wuhan Second Ship Design and Research Institute, Wuhan, 430064, China (email: gychappy@126.com).}

\supplementary{Color versions of one or more of the figures in this article are available online at \href{http://ieeexplore.ieee.org}{http://ieeexplore.ieee.org}.}

\markboth{AUTHOR ET AL.}{SHORT ARTICLE TITLE}
\maketitle

\begin{abstract}
Radio-frequency (RF)-based human activity recognition (HAR) provides a contactless and privacy-preserving solution for monitoring human behavior in applications such as astronaut extravehicular activity monitoring, human-autonomy collaborative cockpit, and unmanned aerial vehicle surveillance. However, real-world deployments usually face the challenge of domain knowledge shifts arising from inter-subject variability, heterogeneous physical environments, and unseen activity patterns, resulting in significant performance degradation. To address this issue, we propose DGAR, a domain-generalized activity recognition framework that learns transferable representations without collecting data from the target domain. DGAR integrates instance-adaptive feature modulation with cross-domain distribution alignment to enhance both personalization and generalization. Specifically, it incorporates a squeeze-and-excitation (SE) block to extract salient spatiotemporal features and employs correlation alignment to mitigate inter-domain discrepancies. Extensive experiments on public RF-based datasets—HUST-HAR, Lab-LFM, and Office-LFM—demonstrate that DGAR consistently outperforms state-of-the-art baselines, achieving up to a 5.81\% improvement in weighted F1-score. The empirical results substantiate the generalization capability of DGAR in real-time RF sensing across dynamic scenarios.
\end{abstract}


\section{INTRODUCTION}
Human activity recognition (HAR) has become essential for intelligent sensing applications, supporting critical tasks such as gesture-based cockpit control, wearable electronics, fall detection, and digital home~\cite{wang2024mobilenet,huang2022channel,rahman2022physics,bianchi2019iot}. Among various sensing paradigms, radio-frequency (RF)-based HAR stands out due to its contactless nature, robustness in low-light or occluded environments, and strong privacy-preserving characteristics~\cite{wang2015review,nirmal2021deep,rahman2022physics}. By analyzing wireless signal variations induced by human movements, RF-based systems can unobtrusively infer physical activities without requiring wearable sensors or cameras, making them particularly suitable for continuous monitoring.

Despite these advantages, deploying RF-based HAR systems in real-world scenarios is challenging. Recognition performance often degrades if the systems encounter variability across users, environmental contexts, or device configurations. These variations induce significant domain knowledge shifts, where the data distributions at test time differ substantially from those seen during training. Such domain discrepancies present substantial obstacles to achieving robust and generalizable recognition.

\begin{figure}[htb]
    \centerline{\includegraphics[width=0.9\columnwidth]{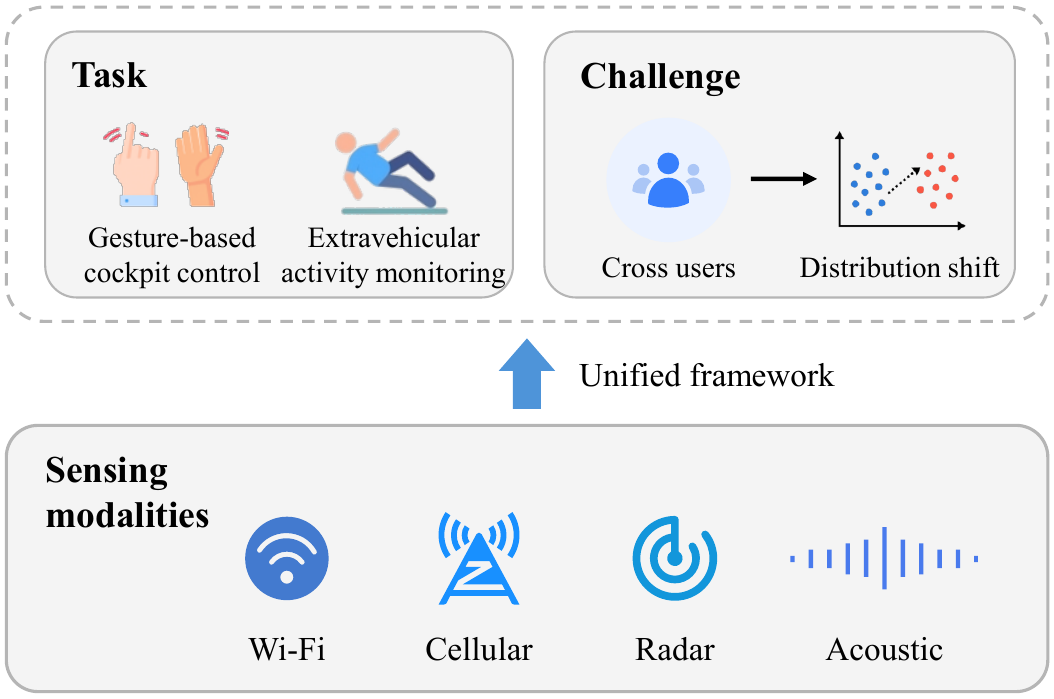}}
    \caption{Conceptual motivation for the proposed framework. The interplay between HAR task requirements and domain shift challenges (e.g., user heterogeneity, environmental dynamics) underscores the necessity for robust and generalizable modeling across diverse sensing modalities, including Wi-Fi, cellular, radar, and acoustic systems.} 
    \label{Intro}
\end{figure}

Fig.~\ref{Intro} summarizes the conceptual motivation of the proposed framework. Accurate and robust HAR requires addressing domain variability and accommodating diverse sensing conditions. Channel state information (CSI), derived from commodity Wi-Fi~\cite{wang2017device} or ultra-wideband (UWB) systems~\cite{chen2021rf}, offers fine-grained multipath measurements capable of recognizing subtle human activities with high precision~\cite{wang2018wi,ma2019wifi,liu2019wireless}. CSI has shown great potential in aerospace sensing applications, including pilot intent detection, human-robot interaction in spacecraft cabins, and unmanned aerial vehicle (UAV)-based through-wall or remote activity recognition in complex terrains.

Driven by the emergence of integrated sensing and communication as a core function in 6G networks~\cite{cui2021integrating,dong2024sensing}, RF-based HAR is expanding beyond Wi-Fi and UWB to encompass modalities such as radar, cellular, and acoustic sensing. Although these modalities vary significantly in waveform characteristics and hardware implementations, they face common challenges regarding robust and transferable representations learning under domain shift conditions.

In recent years, machine learning techniques have significantly advanced RF-based HAR performance. Conventional classifiers, such as support vector machines (SVM)\cite{chen2017robust} and random forests\cite{nunes2017human}, typically rely on hand-crafted features extracted from CSI feedback. However, these methods often fail to provide stable performance in dynamically changing and complex scenarios. To address these limitations, deep learning models have emerged as mainstream approaches, owing to their capability for end-to-end feature extraction and hierarchical representation learning. Convolutional neural networks (CNNs)\cite{zhang2020human} effectively capture local spatial variations in CSI feedback, while recurrent models such as long short-term memory (LSTM) networks\cite{zhang2020data} model temporal dependencies. More recently, attention-based architectures, such as Transformers~\cite{li2021two}, have been utilized to capture long-range dependencies across temporal and spatial dimensions. Furthermore, hybrid models and innovative architectural designs~\cite{li2024harmamba} have further enhanced recognition accuracy under varied motion patterns.

Despite the aforementioned benefits, the generalization capability of existing deep learning models across different users and environments remains limited. This issue is particularly pronounced in RF-based HAR due to the sensitivity of wireless signals to variations in body dynamics, movement styles, and environmental contexts~\cite{wang2022airfi,miao2024wi}. Consequently, inconsistent CSI distributions arise across different deployment domains, severely hindering model performance when applied to previously unseen users or scenarios.

\begin{figure*}[tb]
    \centerline{\includegraphics[width=2.0\columnwidth]{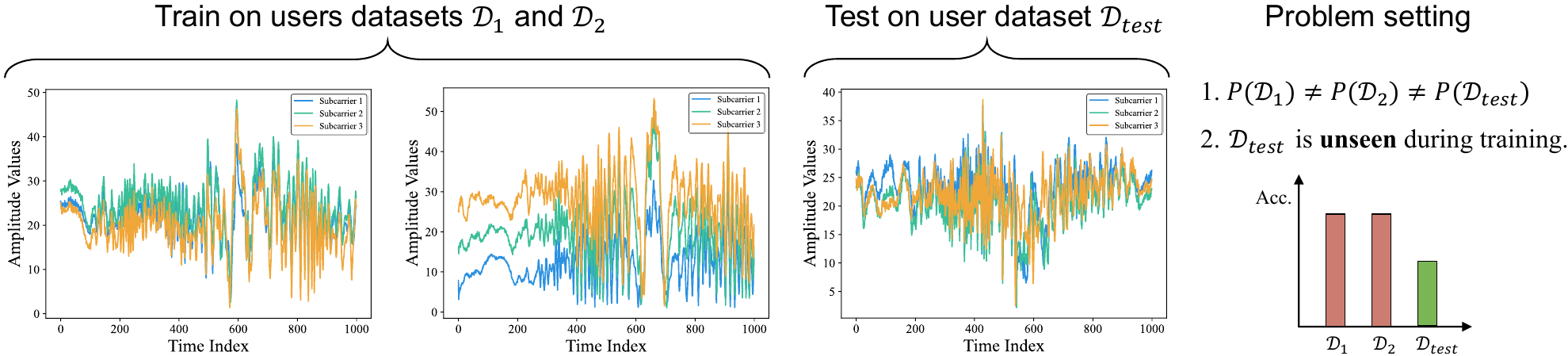}}
    \caption{Variation in CSI distributions among different users. Existing HAR methods suffer substantial performance degradation when test data distributions significantly differ from training domains and remain inaccessible during training.} 
    \label{problemsetting}
\end{figure*}

Fig.~\ref{problemsetting} illustrates this challenge using data from the HUST-HAR dataset~\cite{liu2024tf}, showing distinct CSI patterns generated by different users under identical hardware conditions. This example exemplifies the domain shift problem, where training data distributions significantly diverge from the unseen test domain, i.e., $P(\mathcal{D}_1) \ne P(\mathcal{D}_2) \ne P(\mathcal{D}_{\text{test}})$. Consequently, models trained under independent and identically distributed (i.i.d.) assumptions typically perform poorly in realistic deployment scenarios~\cite{chang2020systematic,soleimani2021cross}.

In order to address the domain knowledge shift problem, researchers have explored transfer learning techniques~\cite{yang2022deep,tang2022learning}. These methods aim to improve model generalization by leveraging auxiliary data from related domains, often through strategies such as fine-tuning on labeled target samples or aligning source and target feature distributions via adversarial learning~\cite{ganin2016domain,long2015learning}. However, most transfer learning approaches require access to target-domain data—either labeled or unlabeled—during training. This assumption is often unrealistic in real-world applications, particularly in privacy-sensitive or mission-critical scenarios.

% This limitation motivates the development of more generalizable frameworks capable of zero-shot deployment, where models must generalize to entirely unseen domains without any prior knowledge of the target data. In this context, domain generalization (DG) has emerged as a promising solution~\cite{wang2022generalizing,qin2022domain}, which seeks to develop models that perform robustly in entirely unseen environments~\cite{li2018domain}. While DG has shown encouraging results in domains such as image recognition and natural language processing~\cite{zhou2022domain,ding2022domain}, its application to RF-based HAR poses distinct challenges. These include the high dimensionality and complex structure of RF data, the sensitivity of wireless signals to environmental dynamics, and the subtle perturbations induced by human movement, all of which lead to significant distribution shifts across deployments.

Motivated by this critical gap, domain generalization (DG) has emerged as a promising strategy~\cite{wang2022generalizing,qin2022domain}, aiming to develop models capable of generalizing effectively to completely unseen environments without prior exposure. While DG techniques have demonstrated encouraging performance in domains like image recognition and natural language processing~\cite{zhou2022domain,ding2022domain}, RF-based HAR presents distinct challenges due to its high-dimensional and complex data structures, environmental sensitivity, and subtle human-induced signal perturbations.

% In this paper, we propose domain-generalized activity recognition (DGAR), a novel framework for zero-shot deployment of RF-based HAR systems. DGAR is grounded in a dual-path strategy: the first is it enhances generalization by aligning shared patterns across domains, and the second is it preserves discriminative power by capturing localized variations specific to individual inputs. Instead of treating all data uniformly, DGAR introduces a flexible modulation mechanism that adapts feature representations per instance, while enforcing statistical consistency across multiple domains. DGAR mainly comprises two core components: (1) Instance-refined feature adaptation, which employs multiple parallel adapters to explore diverse refinements of shared features. A dynamic attention mechanism aggregates these refinements for each input, promoting specialization and adaptability without relying on domain labels. (2) Context-shared feature alignment, which aligns the covariance structures of deep features across all source domains using second-order statistics. This encourages distributional coherence and supports robust generalization to unseen target domains.

In this paper, we propose domain-generalized activity recognition (DGAR), a novel domain generalization framework for zero-shot deployment of RF-based HAR systems. DGAR adopts a dual-path strategy: one branch aligns shared patterns across domains to enhance generalization, while the other captures instance-specific variations to preserve discriminative power. Instead of treating all inputs uniformly, DGAR introduces a flexible modulation mechanism that adapts feature representations per instance, while enforcing statistical consistency across domains. 

The framework comprises two key components:
\begin{itemize}
    \item [(1)] Instance-refined feature adaptation, which employs multiple parallel adapters to explore diverse refinements of shared features. A dynamic attention mechanism aggregates these outputs to promote input-specific specialization without relying on domain labels.
    \item [(2)] Context-shared feature alignment, which minimizes inter-domain covariance discrepancies using second-order statistics, ensuring feature coherence across source domains.
\end{itemize}

% These components are integrated into a unified deep neural network that facilitates resilient learning under domain shifts. To further improve generalization, we incorporate a squeeze-and-excitation (SE) block into the feature extraction backbone, enabling adaptive emphasis on informative subcarriers and spatial channels. Extensive experiments on benchmark RF-based HAR datasets—HUST-HAR, Lab-LFM, and Office-LFM—demonstrate the superior performance of DGAR, with consistent improvements in F1-score over state-of-the-art domain generalization baselines.

These components are integrated into a unified deep neural network. A squeeze-and-excitation (SE) block is further embedded into the feature extractor to emphasize informative subcarriers and spatial channels. Experiments on three RF-based HAR benchmarks—HUST-HAR, Lab-LFM, and Office-LFM—demonstrate DGAR's superiority over state-of-the-art baselines in terms of F1-score and cross-user generalization.

The main contributions are as follows:

\begin{itemize} 
%   \item We propose DGAR, a domain generalization framework tailored for RF-based HAR in dynamically changing human-environment contexts. DGAR addresses the core challenges of non-i.i.d. distributions across users and scenes, without requiring any target domain access. 
  \item We propose DGAR, a domain generalization framework tailored for RF-based HAR under dynamic user and environmental variations, without requiring any target-domain access.
  
%   \item A modular feature refinement mechanism is introduced, where instance-specific adapters generate diverse representations that are fused adaptively. This structure encourages feature specialization while maintaining flexibility to adapt to unseen inputs.
  \item We introduce a modular instance-adaptive refinement mechanism, where parallel adapters generate diverse representations that are adaptively fused via attention.

%   \item We design a cross-domain alignment module that minimizes covariance discrepancies via correlation alignment-based regularization, improving stability and reducing overfitting to training-specific conditions.
  \item We design a cross-domain alignment module that leverages correlation alignment to minimize covariance shifts, improving robustness and reducing overfitting.

%   \item Comprehensive experiments across three representative RF-based HAR datasets are conducted. DGAR consistently achieves state-of-the-art performance in terms of accuracy, F1-score, and cross-person generalization, validating its effectiveness for real-world deployment.
  \item Extensive evaluations across three public datasets validate DGAR's effectiveness, achieving consistent gains in accuracy, F1-score, and domain-level generalization.
\end{itemize}

The rest of this paper is organized as follows. Section~\ref{sec:background} reviews background and related work. Section~\ref{sec:method} describes the proposed DGAR. Extensive experimental evaluations are provided in Section~\ref{sec:experiment}, and Section~\ref{sec:conclusion} concludes the paper.

\section{Background and Related Work}\label{sec:background}
\subsection{RF Channel Modeling}
RF sensing methods typically exploit CSI to distinguish human activities or gestures. In this subsection, we first model the RF channel, then discuss two commonly used RF signal types: Wi-Fi and linear frequency modulation (LFM).

According to~\cite{tse2005fundamentals}, in an indoor multipath environment with $P$ propagation paths, the baseband RF channel model for a transmitter-receiver pair at a carrier frequency $f_c$ can be expressed as
\begin{equation}
    h(t) = \sum_{p=1}^{P} \alpha_p e^{-j2\pi f_c \tau_p} + n(t),
\end{equation}
where $\alpha_p$ is the amplitude of the $p$-th path, and $n(t)$ is Gaussian noise. Additionally, $\tau_p = \tau_p^S + \tau_p^D$, where $\tau_p^S$ and $\tau_p^D$ denote time delays caused by static and dynamic reflections, respectively. For a transmitted signal $s(t)$, the received signal is $y(t) = h(t) \ast s(t)$, where $\ast$ denotes convolution. For simplicity, subsequent derivations focus on a single propagation path (omitting the subscript $p$) and ignore the noise term.

\textbf{Wi-Fi Radio.} Wi-Fi communication systems typically employ orthogonal frequency division multiplexing (OFDM) to distribute digital information across $N$ distinct subcarriers. Let $s_n$ denote the baseband transmitted signal on the $n$-th subcarrier. Under the narrowband assumption, where all subcarriers experience approximately the same delay $\tau$, the received signal for the $n$-th subcarrier is given by~\cite{tse2005fundamentals}
\begin{equation}
    y_n^W = \alpha_n e^{-j 2\pi f_n \tau} s_n, \quad n \in \{1,\dots,N\},
\end{equation}
where $\alpha_n$ and $f_n$ denote the amplitude and frequency of the $n$-th subcarrier, respectively, and $\tau$ represents the effective path delay. The complex channel state $\hat{h}_n$ for each subcarrier can be estimated as $\hat{h}_n = y_n^W / s_n$.

\begin{figure}[tb]
    \centering
    \subfloat[Wi-Fi matrix]{
    \label{heatmapswifi}
    \includegraphics[width=0.45\columnwidth]{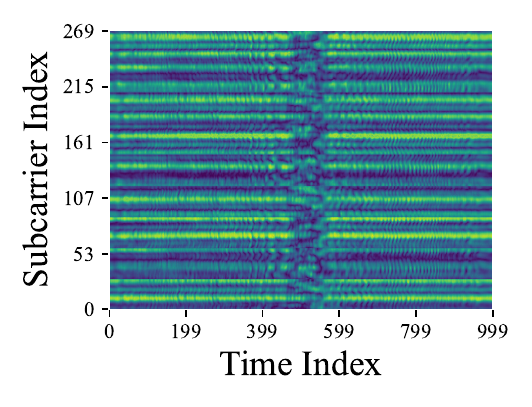}
    }
    \subfloat[LFM matrix]{
    \label{heatmapslfm}
    \includegraphics[width=0.45\columnwidth]{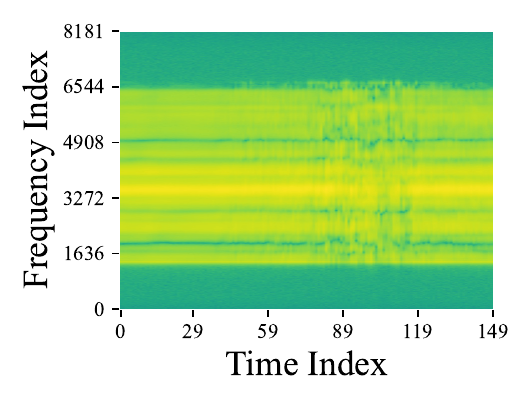}
    }
    \caption{Heatmaps for two RF techniques representing the `walking' activity.}
    \label{heatmaps}
\end{figure}

When multiple OFDM packets are collected over time, the CSI estimates form a two-dimensional matrix of size $N \times T$, where $N$ is the number of subcarriers and $T$ is the number of sequentially received packets. This CSI matrix captures both amplitude and phase information over time, making it valuable for RF sensing applications. Fig.~\ref{heatmaps}~\subref{heatmapswifi} shows an example of a Wi-Fi matrix.

\textbf{LFM Radio.} LFM systems differ from OFDM by continuously sweeping over a bandwidth $B$ within a predefined interval $T_S$, with a sweep rate $\beta = B / T_S$. The transmitted LFM signal can be written as
\begin{equation}
    x^L(t) = e^{j2\pi\left(f_c t + \frac{\beta t^2}{2}\right)}, \quad 0 \le t \le T_S,
\end{equation}
where $f_c$ is the starting frequency of the sweep. After propagating through the channel with a delay $\tau$, the received signal becomes
\begin{equation}
    y^L(t) = \alpha e^{-j2\pi\left(f_c(t-\tau) + \frac{\beta(t-\tau)^2}{2}\right)} \,\Pi(t-\tau),
\end{equation}
where $\alpha$ denotes the channel gain and $\Pi(\cdot)$ is a rectangular window capturing the active chirp duration ($0 \le t-\tau \le T_S$)~\cite{stove1992linear}. In practice, the received signal is sampled and processed using fast Fourier transform (FFT)-based methods to extract frequency-domain features~\cite{chen2006micro}.

When $T$ consecutive `snapshots' are collected, each snapshot is processed into $N$ frequency bins (or chirp segments). These measurements form an $N \times T$ matrix, reflecting the LFM channel's frequency and time variations, as shown in Fig.~\ref{heatmaps}~\subref{heatmapslfm}.
Despite differences in how $N$ and $T$ are interpreted in OFDM (subcarrier indices and packets) versus LFM (frequency bins and snapshots), both methods yield rich 2D data structures. These matrices can be analyzed with signal-processing or learning-based methods to extract micro-motion features, recognize gestures, or perform activity recognition.

Although the above discussion focuses on Wi-Fi and LFM-based RF systems, the underlying modeling approach can be extended to other wave-based sensing modalities such as radar and acoustic sensing. Radar systems, particularly those using frequency-modulated continuous wave (FMCW), emit chirp signals similar to LFM and process the reflected signals via FFT-based methods to estimate target range and velocity~\cite{ding2021radar,yao2021radar}. Likewise, acoustic and ultrasonic systems transmit modulated sound waves that undergo multipath propagation and Doppler shifts when interacting with human motion~\cite{do2021soham,shibata2023listening}. In these cases, the received signals can also be processed into two-dimensional time-frequency representations, forming matrices analogous to CSI or LFM spectrograms. Therefore, the signal modeling and matrix-based representation introduced in this subsection provide a unified foundation for multimodal sensing tasks involving electromagnetic or acoustic wave propagation.

\subsection{RF Sensing Meets Machine Learning}
\begin{figure}[tbp]
    \centering
    \subfloat[Picking up]{
    \includegraphics[width=0.45\columnwidth]{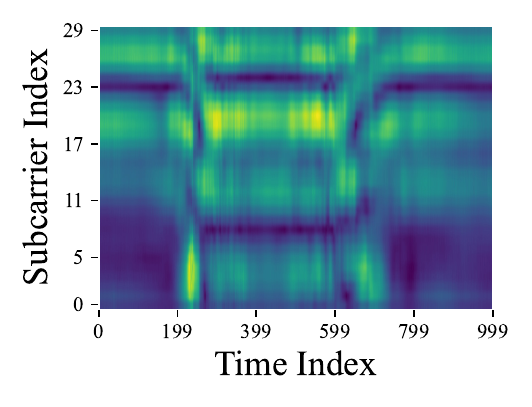}
    }
    \subfloat[Walking]{
    \includegraphics[width=0.45\columnwidth]{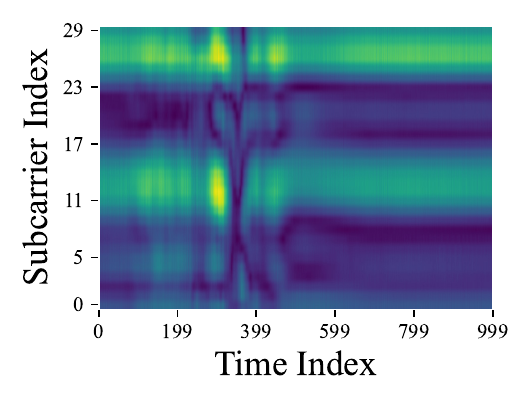}
    }
    \caption{Wi-Fi CSI heatmaps of two different activities: picking up and walking. Recognizing these activities intuitively is nearly impossible for human observers.}
    \label{wificsi}
\end{figure}

Recent advancements in RF-based HAR leverage fine-grained CSI to capture detailed spatiotemporal motion patterns. Commercial Wi-Fi devices, such as Intel 5300 network interface cards~\cite{halperin2011tool}, and specialized RF sensing systems~\cite{taylor2021wireless} facilitate the extraction of CSI, providing richer representations of human activities compared to traditional sensing methods. However, as shown in Fig.~\ref{wificsi}, distinguishing subtle differences—such as picking up an object versus walking—remains challenging for human observers due to the complexity and variability of CSI heatmaps.

To address this challenge, modern machine learning (ML) and deep learning (DL) techniques have been adopted to automatically extract discriminative features from CSI data. These methods effectively leverage both amplitude and phase information, enabling robust activity recognition in complex environments. Traditional ML approaches rely on handcrafted features—such as statistical descriptors, wavelet transforms, or Doppler signatures~\cite{kim2009human,tian2020robust,yang2024generalizable}—which are often based on domain expertise. In contrast, DL models, including CNNs, recurrent neural networks (RNNs), and Transformers, are capable of learning hierarchical spatiotemporal representations directly from raw or minimally preprocessed RF signals~\cite{zhang2020human,zhang2020data,li2021two,ding2019wifi}. This transition reduces the reliance on manual feature engineering while improving recognition performance.

Despite these advances, most existing approaches assume that training and testing data are drawn from the same underlying distribution. This assumption often leads to significant performance degradation when models are deployed in real-world scenarios characterized by unseen users, environments, or deployment configurations.

\subsection{Transfer Learning and Domain Adaptation}
Domain shift poses a major challenge in RF-based HAR, arising from diverse factors such as user characteristics, environmental conditions, hardware configurations, and deployment scenarios. Models trained on a single source domain often suffer performance degradation when deployed in novel target domains. To address this issue, transfer learning and domain adaptation (DA) have emerged as key strategies.

Transfer learning utilizes a model pre-trained on a source domain and fine-tunes it on a target domain with limited labeled data, thereby reducing the need for retraining from scratch. Advanced techniques—such as teacher-student frameworks, self-distillation, and few-shot learning—further enhance cross-domain performance~\cite{deng2023lhar}. For example, Thukral \textit{et al.}~\cite{thukral2023cross} proposed a few-shot learning framework that leverages limited labeled data in the target domain, combined with self-supervised learning and data augmentation, to mitigate domain gaps. Similarly, Zhao \textit{et al.}~\cite{zhao2022domain} curated a large-scale correlated dataset by merging multiple domains and categorizing activities into basic and complex types, enabling the learning of more generalizable features during pre-training.

Domain adaptation, by contrast, extends transfer learning to scenarios with unlabeled or sparsely labeled target-domain data. DA methods—such as adversarial learning and distribution alignment—aim to reduce the discrepancy between source and target distributions. Zhou \textit{et al.}~\cite{zhou2020xhar} proposed XHAR, a domain adaptation framework combining CNNs and bidirectional gated recurrent units (BiGRUs) for spatiotemporal feature extraction, with multiple domain discriminators to align user- and device-specific distributions. Chen \textit{et al.}~\cite{chen2022dynamic} introduced Dynamic Associate Domain Adaptation (DADA), a semi-supervised Wi-Fi HAR framework that integrates an attention-enhanced DenseNet model and dynamically balances labeled and unlabeled samples from the target domain to ensure robust performance in dynamic environments.

Nevertheless, both transfer learning and domain adaptation assume access to some target-domain data during training—a condition that is often unrealistic in practice. In many real-world HAR scenarios, target domains are entirely unknown and inaccessible during model development. This limitation underscores the importance of domain generalization, which aims to build models capable of generalizing to unseen domains without any target-domain supervision.

\subsection{Domain Generalization}
Domain generalization is an emerging research direction that aims to develop models capable of generalizing to unseen target domains by leveraging multiple, diverse source domains~\cite{wang2022generalizing}. Unlike domain adaptation, which assumes partial access to target-domain data during training, DG operates under the stricter assumption that no target-domain data is available. This setting is particularly relevant for real-world HAR tasks, where models must function across novel, heterogeneous environments, user populations, and deployment scenarios.

A core challenge in DG lies in learning a predictive function that remains robust under domain shifts. Recent work has explored various strategies to improve feature invariance and enhance cross-domain alignment. For example, Qin \textit{et al.}~\cite{qin2022domain} proposed Adaptive Feature Fusion for Activity Recognition (AFFAR), a DG framework that dynamically integrates domain-invariant and domain-specific features to improve generalization on public HAR datasets. Yao \textit{et al.}~\cite{yao2024unobtrusive} designed a DG framework for unobtrusive fall detection using radar signals, incorporating domain-specific subclassifiers, entropy regularization, and radar-specific data augmentation to achieve robust generalization across environments and users. Similarly, Liu \textit{et al.}~\cite{liu2025domain} introduced DGSSL, a semi-supervised framework for people-centric activity recognition, combining semi-supervised learning, adversarial training, and reconstruction tasks to enhance domain alignment and predictive consistency on multiple real-world datasets. These efforts underscore the importance of DG in enabling HAR models to adapt to unpredictable, real-world conditions.

Notably, RF signals are characterized by multipath fading, device-specific variations, and modality-dependent noise, making robust modeling substantially more difficult than in vision or wearable-sensor-based settings. While our design is inspired by the adaptive fusion principles explored in AFFAR, DGAR is specifically tailored to handle the unique spatiotemporal complexities of RF data under domain shift. By coupling input-specific feature refinement with global distribution alignment—without relying on any target-domain supervision—DGAR is able to deliver robust zero-shot generalization across previously unseen RF environments.

\section{Proposed Method}\label{sec:method}
This section provides a detailed description of the proposed framework for domain-generalized activity recognition. We begin by formalizing the problem and highlighting the key challenges. We then present the core components of our model, including the feature extraction backbone, instance-refined adapter modules, and the context-shared alignment mechanism. Finally, we describe how these components are jointly trained to enable robust and adaptive inference in unseen target domains.

\subsection{Problem Definition: Domain-Generalized Activity Recognition}
Let $\mathcal{D}_{\text{train}} = \{(\boldsymbol{x}_i, y_i)\}_{i=1}^n$ denote a labeled training set consisting of $n$ activity instances. Each input sample $\boldsymbol{x}_i \in \mathbb{R}^d$ is a $d$-dimensional signal representation, and $y_i \in \{1,\dots,C\}$ is the corresponding activity label. A conventional activity recognition model aims to learn a function $f: \boldsymbol{x} \rightarrow y$ that minimizes the empirical risk
\begin{equation}
    f^* = \arg \min_f \frac{1}{n} \sum_{i=1}^{n} \ell \bigl(f(\boldsymbol{x}_i), y_i \bigr),
\end{equation}
where $\ell(\cdot)$ denotes a task-specific loss function, such as cross-entropy.

However, models trained under this paradigm often exhibit performance degradation when deployed in novel environments or across different user populations. In wireless HAR scenarios, even minor variations in subject morphology, motion dynamics, or ambient signal propagation can result in significant domain shifts that undermine the learned representations. Since it is impractical to pre-collect data for all real-world conditions, enhancing the model's generalization ability becomes essential.

In contrast to domain adaptation or transfer learning~\cite{ding2020rf,thukral2023cross,zhao2020local}, which assume access to target domain data during training, domain generalization operates under a stricter setting: no data from the target domain are available. Instead, we are provided with $K$ related but statistically diverse source domains
\begin{equation}
    \mathcal{D}_{\text{train}} = \{\mathcal{D}_1, \mathcal{D}_2, \dots, \mathcal{D}_K \}, \quad \mathcal{D}_k = \{ (\boldsymbol{x}_i^k, y_i^k) \}_{i=1}^{n_k}.
\end{equation}

All domains share the same input and label spaces, i.e., $\mathcal{X}_{\text{train}} = \mathcal{X}_{\text{test}}$ and $\mathcal{Y}_{\text{train}} = \mathcal{Y}_{\text{test}}$, while their data distributions differ
\begin{equation}
    P_i(\boldsymbol{x}) \neq P_j(\boldsymbol{x}) \neq P_{\text{test}}(\boldsymbol{x}), \quad \forall i \neq j.
\end{equation}

Our goal is to learn a predictive function $f(\cdot)$ that achieves low classification error on an unseen target domain $\mathcal{D}_{\text{test}} = \{ (\boldsymbol{x}_i, y_i) \}_{i=1}^{n_{\text{test}}}$ by leveraging shared structures and latent regularities across the $K$ source domains. This formulation establishes a rigorous benchmark for evaluating the robustness and generalization capabilities of human activity recognition models in real-world deployment scenarios.

\subsection{Main Idea}

\begin{figure*}[tb]
  \centerline{\includegraphics[width=1.60\columnwidth]{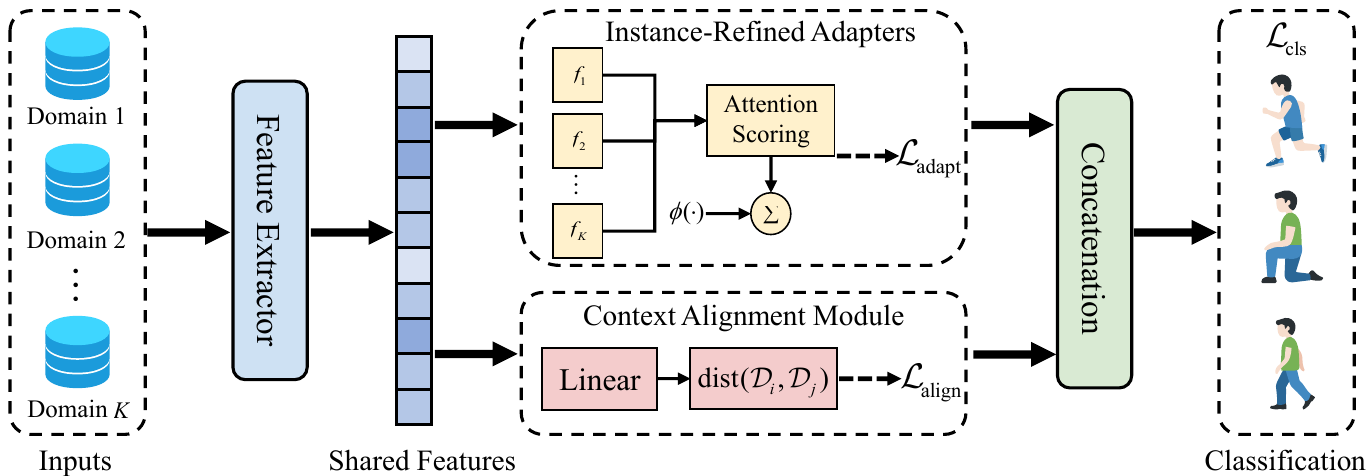}}
  \caption{Block diagram of the proposed DGAR framework. Given input data from $K$ source domains, the model first extracts domain-shared spatiotemporal features via a convolutional backbone. These features are then passed through $K$ parallel instance-refined adapters $\{f_k\}_{k=1}^K$, whose outputs are adaptively fused based on attention scores $\phi(\cdot)$, enabling per-sample specialization. Simultaneously, a context alignment module aligns second-order feature statistics across domains by minimizing pairwise distances $\text{dist}(\mathcal{D}_i,\mathcal{D}_j)$, enforcing domain-invariant structure. The fused representation is then concatenated and fed into a classifier for final activity prediction. The total loss combines classification ($\mathcal{L}_{\text{cls}}$), adapter diversity ($\mathcal{L}_{\text{adapt}}$), and alignment consistency ($\mathcal{L}_{\text{align}}$) objectives.} 
  \label{fig:framework}
\end{figure*}

We propose DGAR, a novel domain generalization framework for RF-based human activity recognition, designed under the practical assumption that no target-domain data are available during training. Despite this constraint, we hypothesize that a previously unseen sample can be effectively approximated by synthesizing transferable features extracted from multiple known training distributions. This hypothesis is grounded in the observation that although RF signal responses (e.g., CSI) may vary significantly across individuals and environments, underlying activity patterns often exhibit common temporal and morphological structures.

To address this, DGAR introduces two complementary mechanisms: (1) a set of instance-refined adapters that capture diverse, input-specific behavioral variations from multiple source domains; and (2) a context-shared feature encoder that learns generalizable representations by aligning the distributions of shared feature activations across datasets.
The overall architecture comprises four key modules: (1) a feature extraction backbone with shared convolutional layers for encoding temporal and frequency-invariant signal patterns; (2) multiple lightweight instance-refined adapters, each modulating the shared features to extract unique behavioral cues; (3) an attention-based fusion mechanism that adaptively aggregates adapter outputs based on their relevance to the input; and (4) a statistical alignment module that minimizes inter-domain representational discrepancies via covariance matching. A visual overview of the framework is provided in Fig.~\ref{fig:framework}.

The feature extraction module, detailed in Table~\ref{tab:feature_extraction}, consists of two residual convolutional blocks with max-pooling and a squeeze-and-excitation block for channel recalibration. A global average pooling layer then produces a compact feature representation, which is routed through $K$ parallel adapters. Each adapter specializes in refining distinct aspects of the representation, and their outputs are fused via a self-attention mechanism to form a composite embedding.

\begin{table}[tb]
    \centering
    \caption{Feature Extraction Module Architecture\protect\footnotemark}
    \label{tab:feature_extraction}
    \setlength{\tabcolsep}{3.5mm}
    \begin{tabular}{ccc}
        \toprule
        \textbf{Layer} & \textbf{Input Shape}  & \textbf{Output Shape} \\
        \midrule
        \multicolumn{3}{c}{\textit{Residual Block 1}} \\
        \midrule
        Conv1 & $B \times D \times L$ & $B \times 128 \times L$ \\
        Conv2 & $B \times 128 \times L$ & $B \times 128 \times L$ \\
        Shortcut & $B \times D \times L$ & $B \times 128 \times L$ \\
        \midrule
        \textbf{MaxPool} & $B \times 128 \times L$ & $B \times 128 \times L/2$ \\
        \midrule
        \multicolumn{3}{c}{\textit{Residual Block 2}} \\
        \midrule
        Conv1 & $B \times 128 \times L/2$ & $B \times 256 \times L/2$ \\
        Conv2 & $B \times 256 \times L/2$ & $B \times 256 \times L/2$ \\
        Shortcut & $B \times 128 \times L/2$ & $B \times 256 \times L/2$ \\
        \midrule
        \textbf{MaxPool} & $B \times 256 \times L/2$ & $B \times 256 \times L/4$  \\
        \midrule
        \textbf{SE Block} & $B \times 256 \times L/4$ & $B \times 256 \times L/4$ \\
        \midrule
        \textbf{Global AvgPool} & $B \times 256 \times L/4$ & $B \times 256$ \\
        \bottomrule
    \end{tabular}
\end{table}
\footnotetext{Each convolutional layer in the residual blocks is followed by a batch normalization (BN) layer and a ReLU activation.}

DGAR is trained by jointly optimizing three objectives
\begin{equation}\label{Eq:loss}
    \mathcal{L} = \mathcal{L}_{\text{cls}} + \lambda \mathcal{L}_{\text{adapt}} + \gamma \mathcal{L}_{\text{align}},
\end{equation}
where $\mathcal{L}_{\text{cls}}$ denotes the classification loss on the fused representations, $\mathcal{L}_{\text{adapt}}$ encourages representational diversity among adapters, and $\mathcal{L}_{\text{align}}$ minimizes distributional discrepancies across source domains. The hyperparameters $\lambda$ and $\gamma$ control the trade-off among these objectives. The following sections elaborate on the construction and optimization of each module.

\subsection{Instance-Refined Feature Adaptation}
To capture diverse variations in input-specific patterns, we introduce a set of instance-refined adapters, each functioning as a lightweight modulation block applied to the shared feature representation. Given an input $\boldsymbol{x}$, the shared encoder $f_e(\cdot)$ first extracts a general representation $\boldsymbol{h} = f_e(\boldsymbol{x})$. This representation is then processed by $K$ parallel adapters $\{f_k\}_{k=1}^K$, resulting in candidate refined features $\boldsymbol{z}_k = f_k(\boldsymbol{h})$.

To aggregate these outputs in a data-adaptive manner, we adopt an attention-based fusion mechanism that assigns importance weights to each adapter output
\begin{equation}\label{Eq:score}
    a_k = \frac{\exp\big(\phi(\boldsymbol{z}_k)\big)}{\sum_{j=1}^{K} \exp\big(\phi(\boldsymbol{z}_j)\big)}, \quad \text{for } k=1,\dots,K,
\end{equation}
where $\phi(\cdot)$ denotes a scoring function (e.g., a shallow MLP) that evaluates the relevance of each refined feature to the given input. The final representation is computed as a weighted sum
\begin{equation}\label{Eq:fusion}
    \boldsymbol{z} = \sum_{k=1}^{K} a_k \cdot \boldsymbol{z}_k.
\end{equation}

This fusion strategy enables the model to dynamically emphasize the most informative refinements without relying on domain labels or explicit routing. Each adapter contributes uniquely to the representation space, supporting fine-grained adaptation to previously unseen inputs.

To promote specialization among adapters and avoid redundancy, we introduce a regularization term that enforces diversity across their outputs
\begin{equation}\label{Eq:adapt}
    \mathcal{L}_{\text{adapt}} = \frac{2}{K(K-1)} \sum_{i<j} \big\| \boldsymbol{\mu}_i - \boldsymbol{\mu}_j \big\|_2^2,
\end{equation}
where $\boldsymbol{\mu}_k$ represents the average output of the $k$-th adapter on its training samples. This regularization encourages representational diversity, ensuring that each adapter captures distinct characteristics.

\subsection{Context-Shared Feature Alignment}

To mitigate feature distribution discrepancies across heterogeneous environments, we introduce a context-shared feature alignment mechanism that enforces consistent representations across all source domains. While the previous module performs input-adaptive refinement, this component ensures global consistency, thereby improving generalization to unseen domains.

Empirically, deeper layers tend to encode more context-specific variations, making them susceptible to source bias~\cite{yosinski2014transferable}. To counteract this, we align the feature distributions from the higher layers across the $K$ source domains $\{ \mathcal{D}_1, \dots, \mathcal{D}_K \}$ using second-order statistics.

We adopt the correlation alignment (CORAL) method~\cite{sun2016deep} to match the covariance structures of feature activations. Let $\boldsymbol{z}_i$ denote the features from domain $\mathcal{D}_i$, with sample covariance matrix $\Sigma_i$. The pairwise distance between two domains is defined as
\begin{equation}
    \text{dist}(\mathcal{D}_i,\mathcal{D}_j) = \left \Vert \Sigma_i - \Sigma_j \right \Vert_F^2,
\end{equation}
where $\left \Vert \cdot \right \Vert_F$ denotes the Frobenius norm. The overall alignment objective is given by
\begin{equation}\label{Eq:align}
    \mathcal{L}_{\text{align}} = \frac{2}{K(K-1)} \sum_{1 \le i < j \le K} \left \Vert \Sigma_i - \Sigma_j \right \Vert_F^2.
\end{equation}

By minimizing $\mathcal{L}_{\text{align}}$, the model learns statistically aligned representations across domains, reducing overfitting to domain-specific covariances and enhancing generalization to unseen conditions.

\subsection{Training and Inference}
Algorithm~\ref{algorithm1} outlines the training procedure for DGAR, which integrates shared feature extraction, instance-adaptive refinement, attention-based fusion, and feature distribution alignment.

\begin{algorithm}[htb]
    \caption{Overall learning procedure of DGAR.}
    \DontPrintSemicolon
    \label{algorithm1}
    \SetKwInOut{Input}{Input}\SetKwInOut{Output}{Output}
    \Input{$K$ labeled training datasets $\mathcal{D}_1, \dots, \mathcal{D}_K$; hyperparameters $\lambda, \gamma$.}
    \Output{Predicted activity labels on the target domain.}
    \BlankLine
    \textbf{Initialize} model parameters $\theta$; \\
    \While{not converged}{
        Sample a mini-batch $\mathcal{B} = \{\mathcal{B}_1,\dots,\mathcal{B}_K\}$ from the $K$ datasets;\\
        Extract shared features $\boldsymbol{h} = f_e(\boldsymbol{x})$;\\
        Compute adapter outputs $\boldsymbol{z}_k = f_k(\boldsymbol{h})$ for $k = 1, \dots, K$;\\
        Compute attention scores $a_k$ in~\eqref{Eq:score} using a learnable scoring function $\phi(\cdot)$;\\
        Fuse features $\boldsymbol{z} = \sum_{k=1}^K a_k \cdot \boldsymbol{z}_k$ in~\eqref{Eq:fusion};\\
        Compute classification loss $\mathcal{L}_{\text{cls}}$ on the fused representations;\\
        Compute adapter diversity loss $\mathcal{L}_{\text{adapt}}$ in~\eqref{Eq:adapt};\\
        Compute alignment loss $\mathcal{L}_{\text{align}}$ in~\eqref{Eq:align} across shared features;\\
        Compute total loss $\mathcal{L} = \mathcal{L}_{\text{cls}} + \lambda \mathcal{L}_{\text{adapt}} + \gamma \mathcal{L}_{\text{align}}$ in~\eqref{Eq:loss};\\
        Update parameters $\theta$ using the Adam optimizer.
    }
    Apply trained model to target samples for prediction.
\end{algorithm}

During training, mini-batches are sampled from the $K$ labeled source domains. The shared encoder extracts general features, which are refined by $K$ instance-specific adapters to generate candidate representations. These are dynamically fused using attention-based weights that assess the relevance of each adapter's output. Simultaneously, a diversity regularization term encourages functional specialization among adapters, while an alignment loss reduces inter-domain feature distribution mismatch.

At inference time, the trained model is fixed. A target sample is processed through the shared encoder and adapters. Attention scores are computed based solely on the adapter outputs, enabling adaptive feature fusion without requiring domain labels. The resulting fused representation is then passed to the activity classifier to generate the final prediction.

\section{Experimental Evaluation}\label{sec:experiment}
This section evaluates the proposed DGAR framework on multiple RF-based activity datasets under domain-generalization settings. We detail the datasets, implementation protocols, baseline methods, and quantitative metrics used for evaluation.

\subsection{Datasets and Preprocessing}
We conduct experiments on three datasets that cover different environments, sensing modalities, and user identities.

\textbf{HUST-HAR}~\cite{liu2024tf} is collected using Intel 5300 Wi-Fi cards operating in a transmitter-receiver configuration. It includes six activities (e.g., lying down, picking up, sitting down, standing, standing up, and walking) performed by six subjects, each repeated 100 times, resulting in 600 samples per activity. We partition the subjects into three groups (two subjects per group), treating each group as one domain.

\textbf{Lab-LFM} is recorded using a USRP-based LFM sensing system in a laboratory setting. Six individuals (one female and five males, aged 20 to 30) performed six activities---kicking, picking up, sitting down, standing, standing up, and walking---each repeated 50 times, for a total of 300 samples per activity (1,800 samples overall). A transmissive RIS is used to enable through-wall sensing. We divide the participants into three two-person domains.

\textbf{Office-LFM} adopts the same LFM setup as Lab-LFM but operates in an office environment with ten subjects (three females, seven males), generating 3,000 samples in total. We group the subjects into five domains (two subjects per domain).\footnote{Both the Lab-LFM and Office-LFM datasets are publicly available at \url{https://github.com/Junshuo-Lau/HUST_HAR_LFM}.}

In all datasets, we simulate domain shifts by assigning each group of individuals as a separate domain. In each run, one domain is held out for testing, and the others are used for training. From the training data, 20\% is further reserved as a validation set for hyperparameter tuning.

\subsection{Baselines and Implementation Details}
We compare DGAR with a series of representative domain generalization methods:

\begin{itemize}
    \item \textbf{Empirical Risk Minimization (ERM)}~\cite{vapnik1991principles}: Trains a single model on the aggregated source domains without explicitly addressing domain discrepancies. This naive baseline minimizes the average empirical loss over all source samples.
    
    \item \textbf{Invariant Risk Minimization (IRM)}~\cite{arjovsky2019invariant}: Encourages the model to learn features that support invariant predictions across domains by enforcing a domain-agnostic optimal classifier. It promotes generalization by discouraging domain-specific shortcuts.
    
    \item \textbf{Domain-Adversarial Neural Network (DANN)}~\cite{ganin2016domain}: Employs an adversarial objective between a feature extractor and a domain discriminator. The feature extractor is trained to fool the discriminator, thereby encouraging feature distributions from different domains to align.
    
    \item \textbf{Group Distributionally Robust Optimization (GroupDRO)}~\cite{sagawa2019distributionally}: Focuses on minimizing the worst-case group loss among all predefined source domains. It adaptively reweights domain contributions during training to improve robustness under domain shifts.

    \item \textbf{Maximum Mean Discrepancy (MMD)}~\cite{gretton2012kernel}: Measures distributional distance in a reproducing kernel Hilbert space (RKHS). By minimizing MMD among source domains, the method encourages the extraction of domain-invariant representations.
\end{itemize}

In addition to the above baselines, we introduce a reference upper bound that assumes access to target-domain data during training:
\begin{itemize}
    \item \textbf{ERM-T}: Serves as an oracle model trained directly on the target domain using an 80\%-20\% train-test split. Although not feasible in practice, it establishes an idealized performance ceiling.
\end{itemize}

To evaluate generalization performance in realistic cross-domain conditions, we adopt a non-IID cross-person HAR setting. Specifically, subjects in each dataset are partitioned into groups, with each group regarded as a separate domain. In each experiment, one domain is designated as the unseen target domain $\mathcal{D}_{\text{test}}$, and the remaining $K$ domains are treated as source domains $\{\mathcal{D}_1,\dots,\mathcal{D}_K\}$. Models are trained exclusively on the source domains, without any access to the target domain during training. To ensure fairness, all models follow the same data splits, training schedules, and backbone architecture.

All methods are implemented in PyTorch 2.2.2 with a unified neural backbone to isolate the effects of algorithmic differences. We use the Adam optimizer with an initial learning rate of $10^{-4}$ for model weights and $10^{-3}$ for the regularization hyperparameters $\lambda$ and $\gamma$. A weight decay of $10^{-5}$ is applied to mitigate overfitting. Each mini-batch contains 32 samples. Training proceeds for up to 100 epochs, with early stopping applied based on validation loss. Learning rate scheduling is handled via the ReduceLROnPlateau strategy, halving the learning rate if the validation performance does not improve for 10 consecutive epochs. All experiments are conducted on an NVIDIA RTX 3090 GPU with 24 GB memory.

We report results using four standard classification metrics: (1) Accuracy, the proportion of correctly classified samples in the test domain; (2) Precision, the weighted average precision across classes; (3) Recall, the weighted average recall across classes; and (4) F1-score, the weighted average F1-score across classes that balances precision and recall.

\subsection{Model Parameter Exploration}
We conduct extensive experiments on the Office-LFM dataset to determine the optimal model configuration prior to benchmarking against other methods. Specifically, we examine three key components of the DGAR framework: backbone architecture, hidden feature dimensionality, and activation function choice.

\begin{figure*}[htb]
    \centering
    \subfloat[]{
    \label{fig:backboneselection}
    \includegraphics[width=0.60\columnwidth]{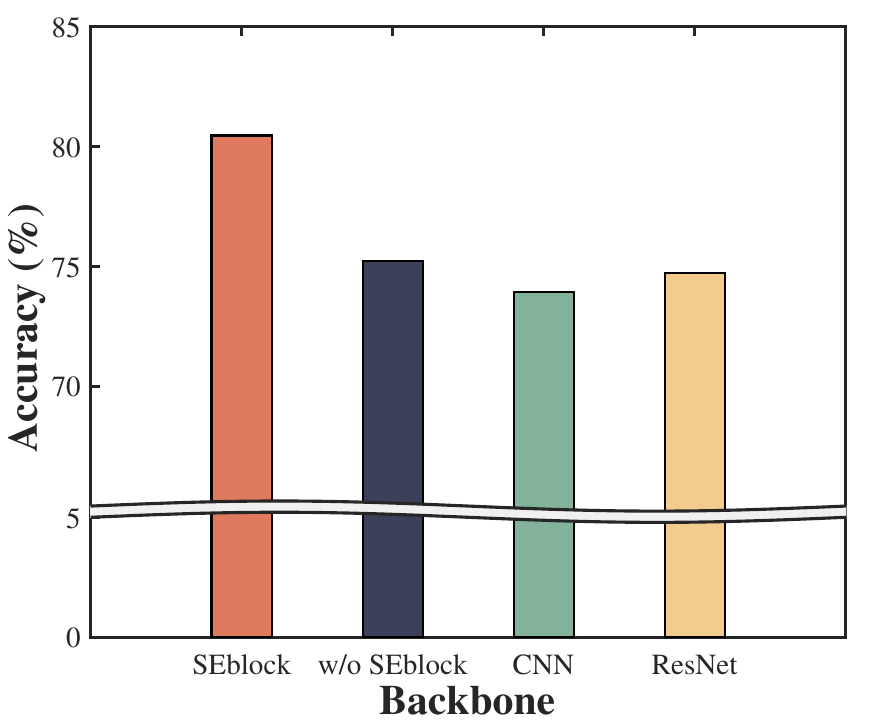}
    }
    \subfloat[]{
    \label{fig:dimensionselection}
    \includegraphics[width=0.60\columnwidth]{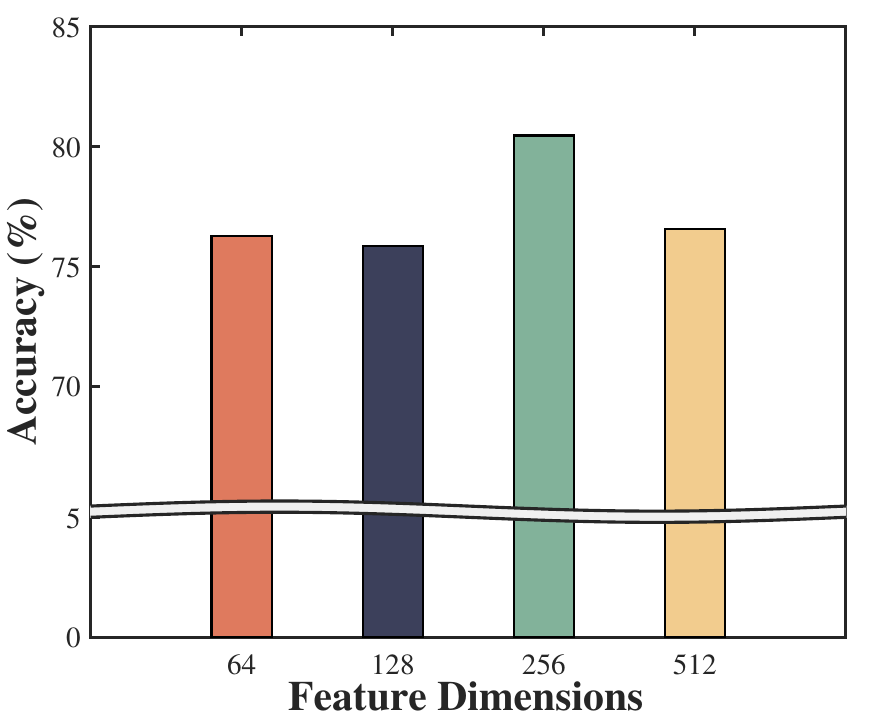}
    }
    \subfloat[]{
    \label{fig:activationselection}
    \includegraphics[width=0.60\columnwidth]{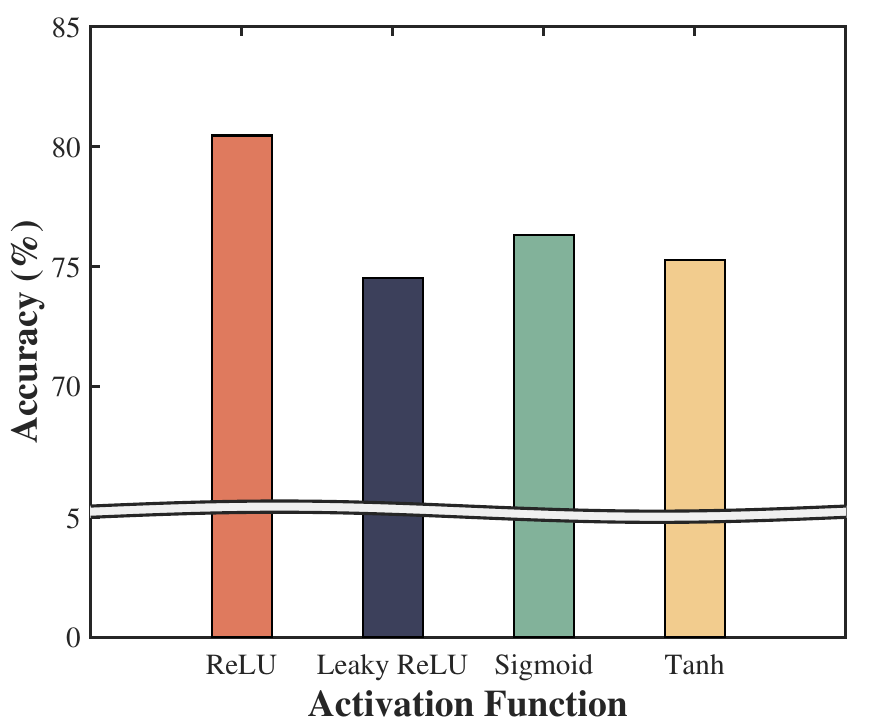}
    }
    \caption{Parameter sensitivity analysis of DGAR. (a) Comparison of different backbones. (b) Impact of the scale of the feature vectors. (c) Classification accuracy against the choice of the activation functions.}
    \label{fig:parameterselection}
\end{figure*}

\subsubsection{Backbone Selection} To extract rich and discriminative representations, we first evaluate the impact of different backbone networks. We compare CNN, ResNet, the proposed model without the SE block, and the proposed model with SE block. The results, shown in Fig.~\ref{fig:parameterselection}~\subref{fig:backboneselection}, demonstrate that incorporating SE blocks consistently yields higher accuracy. This improvement is attributed to the enhanced channel-wise feature recalibration enabled by SE blocks, which amplify informative features while suppressing less useful ones. Based on these observations, we adopt the SE-augmented backbone as the default configuration for subsequent experiments.

\subsubsection{Feature Dimensionality and Activation Function Selection} We then investigate the hidden feature dimensionality, defined as the output size of the global average pooling layer. Specifically, a tensor of shape $B \times 256 \times L/4$ is reduced to $B \times 256$ after pooling. We evaluate dimension sizes in $\{64, 128, 256, 512\}$ to balance representational capacity and computational efficiency. As shown in Fig.~\ref{fig:parameterselection}~\subref{fig:dimensionselection}, a dimension of 256 achieves the highest accuracy, indicating its advantage in retaining discriminative information while remaining computationally efficient.

Finally, we compare several widely used activation functions, including ReLU, Leaky ReLU, Sigmoid, and Tanh (Fig.~\ref{fig:parameterselection}~\subref{fig:activationselection}). Empirical results show that ReLU consistently outperforms the alternatives, likely due to its non-saturating behavior, which facilitates robust feature learning. Therefore, ReLU is adopted throughout the framework.

\subsection{Classification Performance}
We adopt a leave-one-domain-out evaluation protocol. For each dataset, T-$i$ denotes the $i$-th domain held out as the unseen target, while the remaining domains are used for training. Table~\ref{tab:husthar-cls}, \ref{tab:lablfm-cls}, and \ref{tab:officelfm-cls} report the classification performance across all datasets. DGAR consistently achieves the highest weighted F1-scores on HUST-HAR (69.53\%), Lab-LFM (71.68\%), and Office-LFM (80.21\%), surpassing the second-best methods by absolute margins of 2.09\%, 4.52\%, and 5.81\%, respectively. These improvements demonstrate the effectiveness of the proposed dual-branch representation strategy, which integrates both instance-refined and context-shared feature components.

\begin{table*}[htb]
  \centering
  \caption{Weighted F1-Score (\%) on HUST-HAR Dataset. The \textbf{bold} entry indicates the best result except for the ideal condition; the \underline{underlined} entry indicates the second best.}
  \label{tab:husthar-cls}
  \setlength{\tabcolsep}{3.5mm}
  \begin{tabular}{c|cccccc|c}
      \toprule
      Target  & ERM      & IRM        & DANN  & GroupDRO & MMD   & DGAR (ours)    & ERM-T  \\
      \midrule
      T-1     & 69.54    & 62.96      & \underline{69.56} & 68.33    & 66.07 & \textbf{71.30} & 98.34  \\
      T-2     & 71.17    & 69.54      & \underline{71.69} & 69.65    & 71.02 & \textbf{74.86} & 98.47  \\
      T-3     & \underline{61.60} & 57.19 & 57.86 & 57.27    & 59.88 & \textbf{62.43} & 97.37  \\
      \midrule
      Average & \underline{67.44} & 63.23 & 66.37 & 65.08    & 65.66 & \textbf{69.53} & 98.06  \\
      \bottomrule
  \end{tabular}
\end{table*}

\begin{table*}[htb]
  \centering
  \caption{Weighted F1-Score (\%) on Lab-LFM Dataset. The \textbf{bold} entry indicates the best result except for the ideal condition; the \underline{underlined} entry indicates the second best.}
  \label{tab:lablfm-cls}
  \setlength{\tabcolsep}{3.5mm}
  \begin{tabular}{c|cccccc|c}
      \toprule
      Target  & ERM   & IRM   & DANN  & GroupDRO & MMD               & DGAR (ours)    & ERM-T  \\
      \midrule
      T-1     & 66.92 & 67.30 & 69.90 & 70.00    & \underline{70.55} & \textbf{76.09} & 98.61  \\
      T-2     & 64.20 & 63.39 & 60.79 & 65.91    & \underline{67.98} & \textbf{72.96} & 97.50  \\
      T-3     & 61.13 & 59.76 & 58.77 & 61.46    & \underline{62.94} & \textbf{66.00} & 96.38  \\
      \midrule
      Average & 64.08 & 63.48 & 63.15 & 65.79    & \underline{67.16} & \textbf{71.68} & 97.50  \\
      \bottomrule
  \end{tabular}
\end{table*}

\begin{table*}[htb]
  \centering
  \caption{Weighted F1-Score (\%) on Office-LFM Dataset. The \textbf{bold} entry indicates the best result except for the ideal condition; the \underline{underlined} entry indicates the second best.}
  \label{tab:officelfm-cls}
  \setlength{\tabcolsep}{3.5mm}
  \begin{tabular}{c|cccccc|c}
      \toprule
      Target  & ERM   & IRM   & DANN  & GroupDRO & MMD               & DGAR (ours)    & ERM-T  \\
      \midrule
      T-1     & 62.22 & 66.64 & 64.22 & 69.00    & \underline{75.49} & \textbf{78.65} & 99.17  \\
      T-2     & 67.83 & 66.39 & 64.35 & 68.22    & \underline{71.38} & \textbf{81.72} & 98.34  \\
      T-3     & 69.69 & 70.71 & 69.04 & 72.19    & \underline{79.48} & \textbf{83.51} & 100.00 \\
      T-4     & 66.82 & 66.87 & 65.29 & 68.17    & \underline{76.59} & \textbf{78.88} & 98.33  \\
      T-5     & 61.32 & 65.51 & 55.53 & 65.18    & \underline{69.06} & \textbf{78.31} & 95.84  \\
      \midrule
      Average & 65.58 & 67.22 & 63.69 & 68.55    & \underline{74.40} & \textbf{80.21} & 98.34  \\
      \bottomrule
  \end{tabular}
\end{table*}

While methods such as DANN enforce domain-invariant representations through adversarial training, their inability to model input-specific variations limits their generalization capacity. In contrast, DGAR's attention-based fusion of adapter outputs enables dynamic adjustment of representations based on input characteristics, enhancing adaptability across domains. Similarly, IRM and GroupDRO—designed to promote invariant predictors and worst-case robustness, respectively—underperform relative to DGAR, as they insufficiently leverage feature diversity, which is often critical in RF-based HAR.

We further include ERM-T as an upper-bound baseline. This oracle model, trained directly on the target domain, unsurprisingly outperforms all domain generalization methods, including DGAR. This result underscores the advantage of accessing target-specific signals during training. Nevertheless, DGAR achieves competitive performance without any target data, demonstrating its practical utility for real-world scenarios where target-domain access is typically infeasible.
Another noteworthy observation is that average F1-scores on the Office-LFM dataset—comprising five domains—are higher than those on HUST-HAR and Lab-LFM, which include only three each. The increased source diversity exposes the model to a broader range of motion patterns and signal conditions, thereby enhancing the specialization of individual adapters and improving the effectiveness of attention-based fusion. This, in turn, helps DGAR better balance generality and specificity, mitigating overfitting to particular domains and improving robustness on unseen targets.

In summary, the experimental findings yield three key insights: (1) DGAR significantly outperforms existing methods by dynamically fusing multiple refined representations; (2) increasing the number of source domains enhances generalization through diversity-driven learning; and (3) the combination of input-specific adaptation and cross-domain alignment achieves superior robustness to domain shifts. These results support the design choice to move beyond conventional domain-invariant approaches and adopt a more flexible, adaptive representation mechanism tailored for RF-based human activity recognition.

\subsection{Ablation Study}

\subsubsection{Impact of Instance-Refined and Context-Shared Alignment Modules}

\begin{figure*}[htb]
    \centering
    \subfloat[]{
    \label{fig:ablationloss}
    \includegraphics[width=0.65\columnwidth]{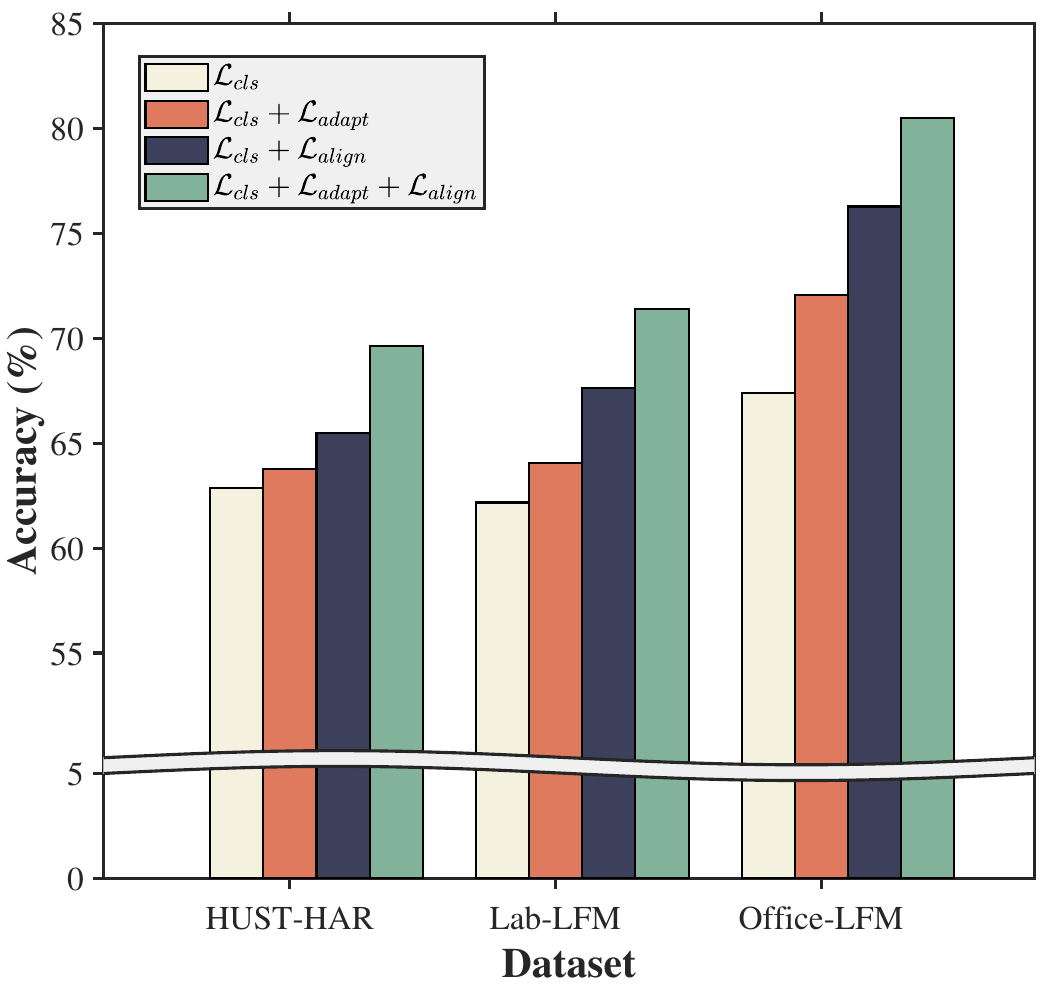}
    }
    \subfloat[]{
    \label{fig:ablationdistance}
    \includegraphics[width=0.65\columnwidth]{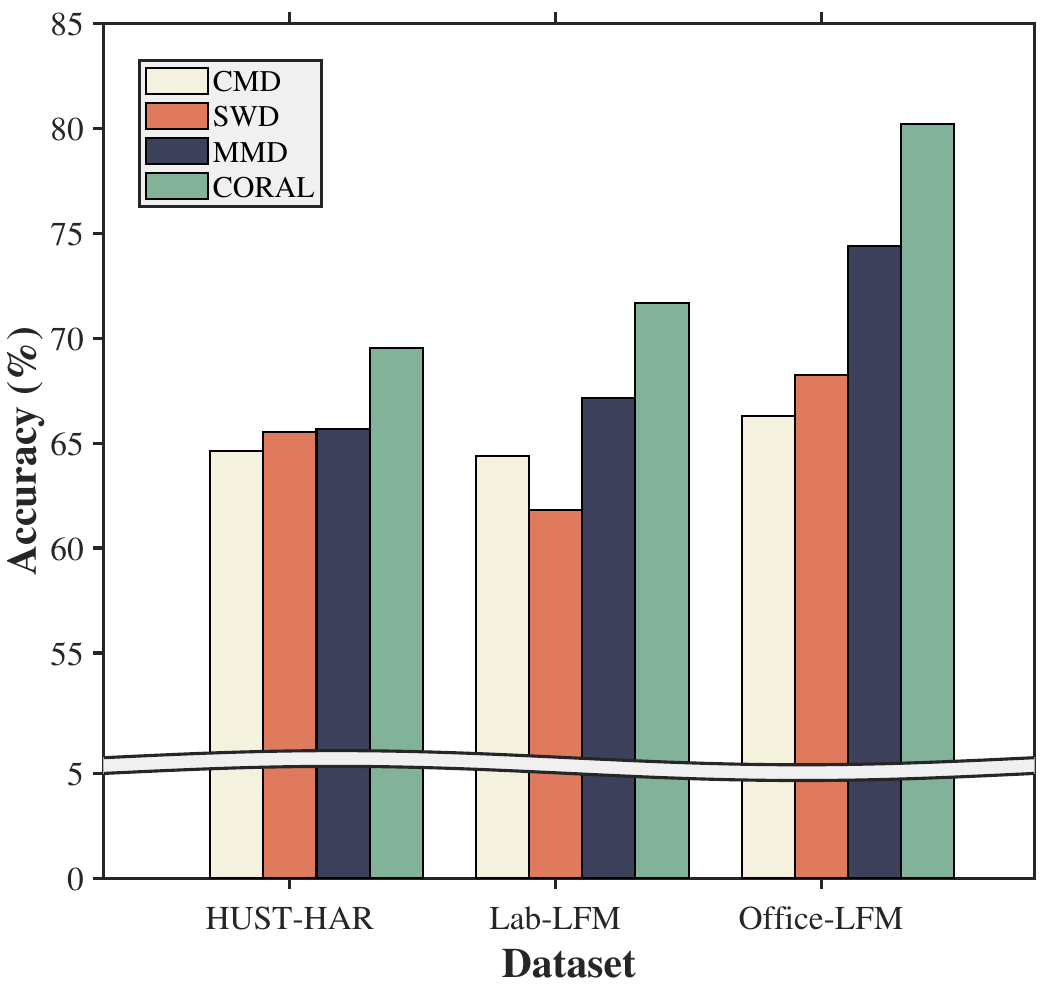}
    }
    \caption{Ablation study results. (a) Effect of instance-refined and context-shared alignment modules. (b) Evaluation of alternative distribution alignment techniques.}
    \label{fig:ablationstudy}
\end{figure*}

To assess the contribution of DGAR's two core components—the instance-refined adapter module and the context-shared alignment module—we conduct an ablation study comparing the following four model variants:

\begin{enumerate}
    \item $\mathcal{L}_{\text{cls}}$: The base model trained with classification loss only;
    \item $\mathcal{L}_{\text{cls}} + \mathcal{L}_{\text{adapt}}$: Adds adapter diversity regularization to enable instance-specific modulation;
    \item $\mathcal{L}_{\text{cls}} + \mathcal{L}_{\text{align}}$: Adds domain alignment loss to enforce feature consistency across source domains;
    \item $\mathcal{L}_{\text{cls}} + \mathcal{L}_{\text{align}} + \mathcal{L}_{\text{adapt}}$ (Full DGAR): Combines both components.
\end{enumerate}

As shown in Fig.~\ref{fig:ablationstudy}~\subref{fig:ablationloss}, each module individually enhances classification accuracy relative to the base model, and their combination yields the best performance across all datasets. These results highlight the complementary nature of global alignment and local adaptation: $\mathcal{L}_{\text{adapt}}$ enables the model to capture subtle input-specific variations, while $\mathcal{L}_{\text{align}}$ ensures consistency across domains.

\subsubsection{Replacing CORAL with Alternative Alignment Losses} To further evaluate the generality of the context-shared alignment module, we replace CORAL with three alternative distribution alignment techniques: MMD, Central Moment Discrepancy (CMD)~\cite{zellinger2017central}, and Sliced Wasserstein Discrepancy (SWD)~\cite{rabin2012wasserstein}.

As illustrated in Fig.~\ref{fig:ablationstudy}~\subref{fig:ablationdistance}, all alignment losses significantly improve performance over the base model, confirming the value of feature-level distribution alignment in DGAR. While CORAL achieves the best performance in most settings, CMD and MMD also produce competitive results. These findings indicate that DGAR is agnostic to the choice of alignment objective and can flexibly incorporate both second-order and higher-order statistics, depending on the data characteristics.

In summary, the ablation study reveals two key insights: (1) integrating instance-refined and context-shared modules improves performance by capturing both domain-invariant and instance-specific features; and (2) the DGAR framework is robust to different alignment strategies, making it broadly applicable to various domain generalization tasks.

\subsection{Hyperparameter Tuning and Inference Efficiency}
To evaluate the robustness of DGAR with respect to its key hyperparameters, we conduct a sensitivity analysis using the Office-LFM dataset, which contains the largest number of source domains. Specifically, we vary $\lambda$ (the weight for the instance-refined adaptation loss) and $\gamma$ (the weight for the context-shared alignment loss), and assess their impact on classification accuracy (\%).

\begin{figure}[htb]
    \centerline{\includegraphics[width=0.95\columnwidth]{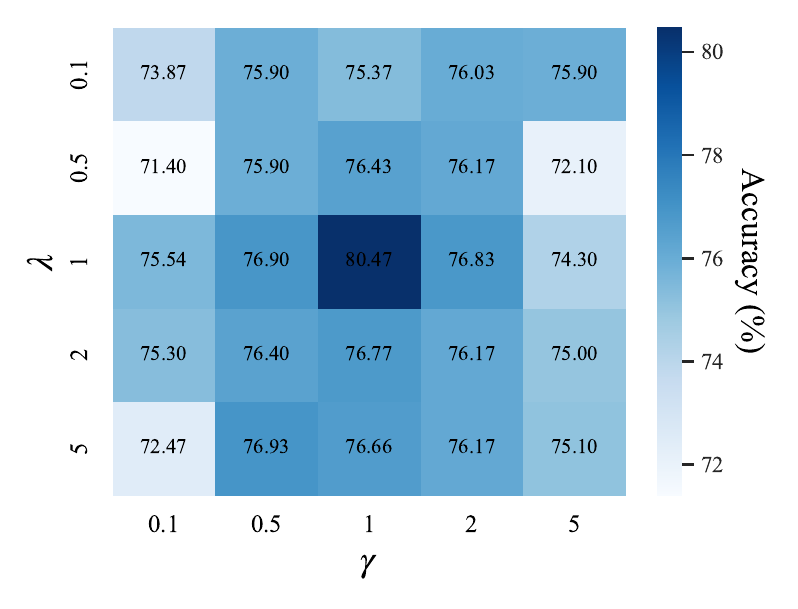}}
    \caption{Sensitivity analysis of $\lambda$ and $\gamma$ on the Office-LFM dataset.} 
    \label{fig:hyperparameter}
\end{figure}

As illustrated in Fig.~\ref{fig:hyperparameter}, model performance degrades noticeably when either hyperparameter is set too low or too high, indicating that balanced regularization between the two components is essential. The highest accuracy of 80.47\% is achieved when both weights are set to 1 (i.e., $\lambda=1$ and $\gamma=1$). Nevertheless, other configurations—such as $\lambda=5$, $\gamma=0.5$ and $\lambda=0.1$, $\gamma=2$—also yield competitive results. These findings confirm that DGAR maintains stable performance across a wide range of hyperparameter values, making it suitable for deployment scenarios where exhaustive tuning is infeasible.

\begin{table*}[tb]
    \centering
    \caption{Inference time, throughput, and accuracy on the Office-LFM dataset.}
    \label{tab:inference}
    \setlength{\tabcolsep}{3.5mm}
    \begin{tabular}{cccccccc}
        \toprule
        Target Model  & ERM   & IRM   & DANN  & GroupDRO & MMD               & DGAR (ours)    \\
        \midrule
        Average time per run (s) & 0.036 & 0.036 & 0.029 & \textbf{0.028} & 0.036 & 0.035 \\
        Throughput (samples/s) & 16793.36 & 16744.80 & 20795.66 & \textbf{21284.65} & 16872.83 & 16977.73 \\
        Accuracy (\%)    & 65.90 & 67.27 & 64.40 & 69.23 & 74.83 & \textbf{80.47} \\
        \bottomrule
    \end{tabular}
\end{table*}

To evaluate the inference efficiency of DGAR, we compare it with all baseline methods on the Office-LFM dataset. As summarized in Table~\ref{tab:inference}, we report the average elapsed time per inference run (averaged over five complete runs of the test set), the corresponding throughput (i.e., the number of samples processed per second), and the final classification accuracy.

As shown in the results, DANN and GroupDRO exhibit slightly faster inference speeds, achieving throughput above 20,000 samples per second. However, their classification accuracy remains below 70\%. In contrast, DGAR offers a competitive inference time (0.035 seconds per run) and throughput (16,977.73 samples/s), while substantially outperforming all baselines in accuracy. This demonstrates that DGAR achieves an effective trade-off between computational speed and recognition precision, enabling high-accuracy performance with near real-time inference capability.

In conclusion, the DGAR framework demonstrates both hyperparameter robustness and computational efficiency. Its ability to maintain superior recognition accuracy without incurring significant latency makes it well-suited for deployment in latency-sensitive human activity recognition applications.

\section{Conclusion}\label{sec:conclusion}
This paper presented DGAR, a domain-generalized framework for RF-based human activity recognition, designed to address the inherent challenges of domain shifts without requiring access to target-domain data. The framework jointly leverages instance-refined adaptation and context-shared alignment mechanisms to capture both input-specific variations and cross-domain invariances. Specifically, DGAR integrates attention-based modulation and feature distribution alignment (via correlation alignment) to ensure robust and transferable representations across heterogeneous environments. Extensive evaluations on three datasets—HUST-HAR, Lab-LFM, and Office-LFM—demonstrate that DGAR consistently outperforms state-of-the-art baselines in terms of weighted F1-score, particularly in cross-subject generalization scenarios. These results validate the effectiveness of DGAR's dual-path modeling strategy and its adaptability to diverse sensing conditions.

Future work will investigate the integration of meta-learning strategies to enable rapid adaptation of DGAR to previously unseen radar deployment scenarios, such as varying flight altitudes or mission-specific motion patterns. We will also extend the DGAR architecture to support multimodal fusion of RF signals and onboard sensor data (e.g., radar, IMU, or thermal imaging) for enhanced activity recognition in aerospace environments. Furthermore, we plan to optimize DGAR for real-time inference on resource-constrained aerial platforms, facilitating robust human or object activity sensing in dynamic outdoor or in-flight conditions.

\bibliographystyle{IEEEtran}
\bibliography{Reference}

\end{document}